\documentclass[10pt,pra,twocolumn,amssymb,nofootinbib,showpacs,letterpaper]{revtex4}

\pdfoutput=1

\usepackage{bm,braket,graphicx,amsmath,bbm}

\usepackage[pdftitle={Noise gates for decoherent quantum circuits},
            pdfauthor={Angelo Bassi, Dirk - André Deckert},
            pdfpagemode=UseOutlines,
            pdfstartview=FitH,
            bookmarks=true,
            bookmarksopen=true,
            bookmarksnumbered=true,
            bookmarkstype=toc,
            colorlinks,
            linkcolor=blue,
            citecolor=blue,
            urlcolor=blue]{hyperref}

\begin{document}

\title{Noise gates for decoherent quantum circuits}

\author{Angelo Bassi}
\email{bassi@ts.infn.it}
\address{Department of Theoretical Physics,
University of Trieste, Strada Costiera 11, 34014 Trieste, Italy.
\\  Mathematisches Institut der LMU, Theresienstr. 39, 80333
M\"unchen, Germany.\\
Istituto Nazionale di Fisica Nucleare, Sezione di Trieste, Strada
Costiera 11, 34014 Trieste, Italy.}
\author{Dirk - Andr\'e Deckert}
\email{deckert@mathematik.uni-muenchen.de}
\affiliation{Mathematisches Institut der LMU, Theresienstr. 39,
80333 M\"unchen, Germany. \\
Department of Theoretical Physics, University of Trieste, Strada
Costiera 11, 34014 Trieste, Italy.}

\begin{abstract}
A major problem in exploiting microscopic systems for developing a
new technology based on the principles of Quantum Information is the
influence of noise which tends to work against the quantum features of
such systems. It becomes then crucial to understand how noise
affects the evolution of quantum circuits: several techniques have
been proposed among which stochastic differential equations (SDEs)
can represent a very convenient tool. We show how SDEs naturally map
any Markovian noise into a linear operator, which we will call a
noise gate, acting on the wave function describing the state of the
circuit, and we will discuss some examples. We shall see that these
gates can be manipulated like any standard quantum gate, thus
simplifying in certain circumstances the task of computing the overall
effect of the noise at each stage of the protocol. This approach
yields equivalent results to those derived from the Lindblad
equation; yet, as we show, it represents a handy and fast tool for performing computations, and moreover, it allows for fast numerical simulations and generalizations to non Markovian noise. In detail we review the depolarizing channel and the generalized amplitude damping channel in terms of this noise gate formalism and show how these techniques can be applied to any quantum circuit.
\end{abstract}
 \maketitle

\section{Introduction}

The development of both theoretical and experimental research on
quantum information is opening the way to a new technology based on
the quantum properties of microscopic systems; its potentialities
are extraordinary, but the possibility of {\it actually}
exploiting quantum properties for building new physical devices is
not yet completely clear. The main reason for this limitation, as is
well known, is that quantum systems are highly sensitive to the
influence of the surrounding environment which tends to destroy
quantum coherence: it becomes important to analyze to what extent
external influences disturb the time evolution of quantum systems.


The most common techniques employed in studying the interaction of a quantum system with an environment mainly rest on the master equation approach~\cite{bp}, the operator-sum representation method~\cite{nc}, and stochastic unravelings of master equations in terms of random quantum jumps~\cite{qja_ns_as} or stochastic differential equations (SDEs)~\cite{sde}. The SDE approach has gained increasing popularity in recent years, but attention has focused mainly on non linear SDEs which in general are difficult to work with. In this paper we show how SDEs can be a flexible and handy mathematical tool for analyzing many physical situations analytically and numerically. The key property of SDEs we will use is that among the different stochastic unravelings of a given master equation of the Lindblad type, there is always one that is {\it linear}~\cite{lsde}: by resorting to this specific unraveling, the power of the superposition principle can be used to analyze the evolution of the open system. As we shall see, the effect of the environment can then be described in terms of a linear and stochastic matrix, which can be manipulated like any other standard quantum gate, when any quantum protocol is analyzed. For this reason we will call it a {\it noise gate}.

This approach has some advantages with respect to the other ones, at
least in certain circumstances:
\begin{enumerate}
\item During computations it allows one to work with {\it state vectors} instead of density matrices, even if the system is open; this makes the analysis
simpler, since it allows the system to be treated as if it were closed,
the effect of the environment being modeled by a random potential.
\item It is predictively equivalent to the Lindblad approach in the
limit of Markovian interactions; at the same time, it can be generalized to non Markovian
dynamics~\cite{nmd}, which are a subject of increasing theoretical
interest~\cite{bp,nmme} for the description of several important
physical phenomena~\cite{appl}.
\item In many important cases it
allows exact mathematical results to be computed; also those, such as
the long-time behavior or the limit for a large number of qubits,
which cannot be computed numerically. When exact results cannot be
obtained, it allows for a perturbation analysis or
for fast numerical simulations~\cite{klo}. Alternatively, in some
cases, part of the analysis can be done analytically and part
numerically, thus simplifying the overall work.
\end{enumerate}
The paper is organized as follows. Section~\ref{sec:one} sets up
the general formalism which allows the effect of the
environment to be expressed through noise gates. In Sec.~\ref{sec:two} we compute the corresponding noise gates for four standard noise channels. Sections~\ref{sec:three}-\ref{sec:five} contain pedagogical examples of how the noise gate formalism can be applied to any quantum circuit. In Sec.~\ref{sec:six} we conclude with final remarks.

\section{Master equations, SDEs and noise gates}
\label{sec:one}

Let us consider a master equation of the Lindblad type:
\begin{equation} \label{eq:master}
\frac{d}{dt}\, \rho(t) = - \frac{i}{\hbar}\, \left[ H, \rho(t)
\right] +  \gamma L \rho(t) L^{\dagger} - \frac{\gamma}{2} \, \{
L^{\dagger} L, \rho(t) \},
\end{equation}
where the Lindblad operator $L$ (for the sake of simplicity, we consider
only one such operator) summarizes the effect of the environment on
the quantum system, and $\gamma$ is a coupling constant. It is well
known that Eq.~\eqref{eq:master} allows for different unravelings
in terms of SDEs; what perhaps is less known is that, among such
unravelings, there is always one which is linear~\cite{lsde},
namely,
\begin{equation} \label{eq:lsd}
d |\psi_{t} \rangle = \left[ -\frac{i}{\hbar}\, H dt + i
\sqrt{\gamma}\, L\, dW_{t}  - \frac{1}{2}\, \gamma L^{\dagger} L dt
\right] |\psi_{t} \rangle,
\end{equation}
where $W_{t}$ is a Brownian motion defined on a probability space
$(\Omega, {\mathcal F}, {\mathbb P})$. When $L$ is a self-adjoint
operator, Eq.~\eqref{eq:lsd} preserves the norm of $|\psi_{t}
\rangle$; however this is no longer true for general operators. In
such cases one can always replace the above equation with a
non linear one, which is norm preserving~\cite{ad}. However,
this is not necessary since, even for the unnormalized state
$|\psi_{t}\rangle$ being a solution of Eq.~(\ref{eq:lsd}), one can easily
prove that
\begin{equation} \label{eq:relgfg}
\rho(t) \; \equiv \; {\mathbb E} [ |\psi_{t}\rangle \langle \psi_{t}
| ],
\end{equation}
which means that, when the stochastic average ${\mathbb E}$ (with
respect to the measure ${\mathbb P}$) is computed, the predictions
of Eq.~\eqref{eq:lsd} are {\it equivalent} to those of
Eq.~\eqref{eq:master}. In this sense Eq.~\eqref{eq:lsd} is an
unraveling of Eq.~\eqref{eq:master}.

Now, because of linearity, the solution of Eq.~\eqref{eq:lsd} can be
generally expressed as follows:
\begin{equation}\label{eq:noisegate}
|\psi_{t} \rangle \; = \; N(t, t_{0}) |\psi_{t_{0}}\rangle,
\end{equation}
and the linear operator $N(t, t_{0})$, which from now on we will
refer to as the {\it noise gate}, can be treated like any other
quantum gate, except for the fact that in general it is not unitary.
Such a gate of course depends on the noise $W_{t}$, which means that
for each realization of the noise the system evolves following
different ``histories''. So all one has to do is to solve
Eq.~\eqref{eq:lsd} to get the correct expression for the noise gate
$N(t, t_{0})$, which can then be inserted in a quantum circuit in
the usual way; after having computed the square modulus of the
probability amplitudes, their stochastic average can be calculated
to obtain the correct physical predictions.

In the following we give the exact expressions of four important
quantum noise gates, and on the basis of two example quantum circuits
we show how the proposed noise gate formalism can be used to treat
the effects of quantum noise in any quantum circuit.

A note before concluding. In the Introduction we mentioned that other (infinitely many different) unravelings of Eq. (\ref{eq:master}) in terms of SDEs are possible \cite{sde}; in such cases Eq. (\ref{eq:lsd}) is replaced by other, structurally different, SDEs, which in general depend in a non linear way on $|\psi_{t}\rangle$. Despite this, the relation (\ref{eq:relgfg}) still holds true for any such equation, implying that, at the statistical level, the predictions computed through the SDEs are equivalent to those computed through the master equation. This also implies that, at the statistical level only, all effects of non linearity are "washed away".  The disadvantage of such unravelings with respect to ours is that, precisely due to their non linearity, the solution of the SDE can not be written as in (\ref{eq:noisegate}) in terms of a linear operator, and therefore the superposition principle can not be used to infer the effect of the noise on a generic state, once its effect on a basis is known.

\section{examples of noise gates}
\label{sec:two}

\subsection{Bit Flip, Phase Flip, Bit-Phase Flip Channels}

These three channels are accurately described in~\cite{nc} within
the framework of the quantum operator-sum representation. It is not
difficult to represent them in terms of a master equation of the
form~\eqref{eq:master}: the corresponding Lindblad operator $L$
turns out to be one of the Pauli matrices, $\sigma_{x}$ for the bit flip channel, $\sigma_{z}$
for the phase flip channel and $\sigma_{y}$ for the bit-phase flip
channel.

For each of these operators, the SDE~\eqref{eq:lsd} becomes
\begin{equation} \label{eq:sdfsd}
d |\psi_{t}\rangle = \left[ i \sqrt{\gamma}\, \sigma_{\kappa} dW_{t} \, -
\, \frac{1}{2} \gamma I dt \right] |\psi_{t}\rangle,
\end{equation}
where $I$ is the identity matrix and $\kappa = x,y,z$. Since the matrices
appearing in Eq. (\ref{eq:sdfsd}) obviously commute, it can
be solved by means of standard techniques~\cite{arn}; the solutions
are:
\begin{eqnarray}
N_{\makebox{\tiny BitFl}}(t, t_{0}) & = & \exp\left[ i \sqrt{\gamma}
\sigma_{x} (W_{t} - W_{t_{0}}) \right], \label{eq:bfc1} \\
N_{\makebox{\tiny PhFl}}(t, t_{0}) & = & \exp\left[ i \sqrt{\gamma}
\sigma_{z} (W_{t} - W_{t_{0}}) \right], \label{eq:bfc2}\\
N_{\makebox{\tiny Bit-PhFl}}(t, t_{0}) & = & \exp\left[ i
\sqrt{\gamma} \sigma_{y} (W_{t} - W_{t_{0}}) \right].\label{eq:bfc3}
\end{eqnarray}
As we see, the above gates lead to a nice physical interpretation of
the effect of the environment on a qubit: it randomly rotates the
qubit along a specific direction, the randomization being
proportional to the strength of the coupling constant $\gamma$.

\subsection{Amplitude Damping Channel}

The amplitude damping channel is also described in~\cite{nc}, and
the associated Lindblad operator is $\sigma^{-} \equiv | 0
\rangle\langle 1 |$; written in terms of the components $\alpha_{t}
\equiv \langle 0 | \psi_{t} \rangle$ and $\beta_{t} \equiv \langle
1| \psi_{t} \rangle$, Eq.~\eqref{eq:lsd} becomes:
\begin{eqnarray}
d \alpha_{t} & = & \sqrt{\gamma} \beta_{t} dW_{t}, \\
d \beta_{t} & = & - (\gamma/2) \beta_{t} dt.
\end{eqnarray}
The solution, expressed in matrix notation, is:
\begin{eqnarray}
N_{\makebox{\tiny AmDa}}(t, t_{0}) & = & \left(
\begin{array}{cc}
1 & i \varphi_{(t, t_{0})} \\
0 & e^{-\frac{\gamma}{2}(t - t_{0})}
\end{array}
\right), \\
\varphi_{(t, t_{0})} & = & \sqrt{\gamma} \int_{t_{0}}^{t}
e^{-\frac{\gamma}{2}s} dW_{s}. \label{eq:bfc4}
\end{eqnarray}
This channel models loss of energy to the environment: the $|1\rangle$
state decays to $|0\rangle$ at a given rate $\gamma$ while $|0\rangle$ is
stable.

We now turn to first applications of this formalism, during which we
will spot some general features of the noise gates, which are useful
for simplifying the calculations.

\section{Application 1: Noisy C-NOT gate}
\label{sec:three}

\begin{figure}[b]
\begin{center}
\includegraphics[scale=1.4]{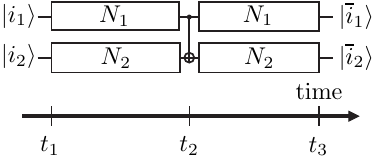}
\caption{CNOT gate for two qubits subject to environmental noise.
We assume that the two noises represented by the gates $N_{1}$ and
$N_{2}$ are independent.} \label{cnot}
\end{center}
\end{figure}

As a first example, we analyze the controlled-NOT gate (CNOT) by assuming that, before and
after its application, the involved pair of qubits are subject to
environmental noise, as shown in Fig.~\ref{cnot}.
Such a quantum circuit is interesting because the final state of the target
qubit will depend not only on the noise acting on it but, through
the CNOT gate, also on the noise acting on the control qubit: the noise gate
formalism allows this dependence to analyzed in simple terms. For
definiteness, let us take $|0,0\rangle$ as the input state. Let us
moreover assume that the noises acting on the two qubits are
independent of each other; this assumption is of course justified only if, e.g.,
the two physical states encoding the two qubits are separated by
more than the correlation length of the noise, so that the
surrounding environment acts independently on them.

The noisy CNOT gate acts on the qubits as follows:
\begin{eqnarray}
|0,0\rangle & \xrightarrow[]{(1)} & N_1(t_2,t_1)\otimes N_2(t_2,t_1)\ket{0,0}\nonumber\\
 & = & \sum_{i,j=0}^{1}
n_{1}^{(i,0)}(t_{2},t_{1}) n_{2}^{(j,0)}(t_{2},t_{1}) |i,j\rangle \nonumber \\
& \xrightarrow[]{(2)} & \sum_{i,j=0}^{1} n_{1}^{(i,0)}(t_{2},t_{1})
n_{2}^{(j,0)}(t_{2},t_{1}) |i,i\oplus j\rangle
\nonumber \\
& \xrightarrow[]{(3)} & \sum_{i,j=0}^{1} \sum_{m,\ell=0}^{1}
n_{1}^{(i,0)}(t_{2},t_{1}) n_{2}^{(j,0)}(t_{2},t_{1}) \nonumber\\
& & \cdot n_{1}^{(m,i)}(t_{3},t_{2}) n_{2}^{(\ell,i\oplus
j)}(t_{3},t_{2}) |m,\ell\rangle,
\end{eqnarray}
where $n_{\alpha}^{(i,j)}(t_{b},t_{a})$ is the $(i,j)$-th
coefficient of the matrix $N_{\alpha}(t_{b},t_{a})$, $\alpha = 1,2$.
Step (1) takes into account the effect of the environment from time
$t_{1}$ to time $t_{2}$, after which the CNOT gate is applied: the latter corresponds to step (2). In step (3) we assume that the environment
acts on the qubits until time $t_{3}$. Let us call
$|\overline{0,0}\rangle$ the final state.

For a closed system we would have of course $|\overline{0,0}\rangle
= |0,0\rangle$, but with the environment interacting with the two
qubits, $|\overline{0,0}\rangle$ becomes a random entangled state;
to understand the effect of the noise, let us compute the fidelity
$F \equiv {\mathbb E} [ | \langle 0, 0 |\overline{0,0}\rangle|^2 ]$
of the noisy protocol (the stochastic average ${\mathbb E}$ takes
into account all possible realizations of the noise). Because of the
Markov property of Wiener processes, we have the following property:

\vskip 0.3cm \noindent \textsc{Property 1.} Two gates $N(t_{2},
t_{1})$ and $N(t_{4}, t_{3})$ are {\it independent} whenever
$(t_{1}, t_{2}) \cap (t_{3}, t_{4})$ is empty. This follows directly
from the fact that a Brownian motion has independent increments.
\vskip 0.3cm

\noindent According to this rule, and keeping in mind the assumption
that the noise gates acting on the two qubits are also independent,
$F$ is a combination only of terms having the form ${\mathbb E} [
n_{\alpha}^{(i,k)}(t_{b},t_{a})
n_{\alpha}^{(i',k')}(t_{b},t_{a})^{\star}]$ with $\alpha = 1,2$ and
$(t_{b},t_{a}) = (t_{2}, t_{1})$ or $(t_{3}, t_{2})$, since all
other terms vanish when averaged, because of the statistical
independence. Now it is just a matter of choosing the type of noise
that best describes the environment, and computing the required
expectation values.

Let us consider, as an example, the {\it bit flip} gate given in
Eq.~\eqref{eq:bfc1}:
\begin{equation}
N_{\makebox{\tiny BitFl}}(t, t_{0}) = \left(
\begin{array}{cc}
\cos( \sqrt{\gamma} \Delta W_{t}) & i \sin (\sqrt{\gamma} \Delta W_{t}) \\
i \sin( \sqrt{\gamma} \Delta W_{t}) & \cos (\sqrt{\gamma} \Delta W_{t})
\end{array}
\right),
\end{equation}
with $\Delta W_{t} = W_{t} - W_{t_{0}}$; one easily verifies that,
for a standard Wiener process, the following equalities hold true:
\begin{eqnarray}
{\mathbb E}[ \cos^2( \sqrt{\gamma_{\alpha}} ( W_{t_{b}} - W_{t_{a}}))]
& = &
p_{\alpha}(t_{b} - t_{a}) \\
& \equiv &  \frac{1 + \exp[-2\gamma_{\alpha}(t_{b} - t_{a})]}{2}
\nonumber \\
{\mathbb E}[ \sin^2(\sqrt{\gamma_{\alpha}} ( W_{t_{b}} - W_{t_{a}}))]
& = & \overline{p}_{\alpha}(t_{b} - t_{a}) \\
& \equiv & 1 - p_{\alpha}(t_{b} - t_{a}), \nonumber
\end{eqnarray}
while ${\mathbb E}[ \cos(\sqrt{\gamma_{\alpha}} ( W_{t_{b}} -
W_{t_{a}})) \sin (\sqrt{\gamma_{\alpha}} ( W_{t_{b}} - W_{t_{a}}))] =
0$. From these expressions one can derive the following expressions
for the correlation functions of the coefficients of the noise gate:
\begin{eqnarray}
{\mathbb E} [ n_{\alpha}^{(i,k)}(t_{b},t_{a})
n_{\alpha}^{(i',k')}(t_{b},t_{a})^{\star}] & = & p_{\alpha}(t_{b} -
t_{a}) \\
& & \text{if $i=k$, $i'=k'$}; \nonumber \\
{\mathbb E} [ n_{\alpha}^{(i,k)}(t_{b},t_{a})
n_{\alpha}^{(i',k')}(t_{b},t_{a})^{\star}] & = &
\overline{p}_{\alpha}(t_{b} - t_{a}) \\
& & \text{if $i\neq k$, $i' \neq k'$}; \nonumber \\
{\mathbb E} [ n_{\alpha}^{(i,k)}(t_{b},t_{a})
n_{\alpha}^{(i',k')}(t_{b},t_{a})^{\star}] & = & 0 \\
& & \text{in all other cases}. \nonumber
\end{eqnarray}
Given this, one easily gets the following expression for the
fidelity $F$ as a function of the coupling constants
$\gamma_{\alpha}$ and of the time intervals during which the noises
act on the CNOT gate:
\begin{eqnarray}
F & = & p_{1}(t_{2} - t_{1})p_{2}(t_{2} -
t_{1})p_{1}(t_{3} - t_{2})p_{2}(t_{3} - t_{2}) \nonumber \\
& + & \overline{p}_{1}(t_{2} - t_{1})p_{2}(t_{2} -
t_{1})\overline{p}_{1}(t_{3} -
t_{2})\overline{p}_{2}(t_{3} - t_{2}) \nonumber \\
& + & p_{1}(t_{2} - t_{1})\overline{p}_{2}(t_{2} - t_{1})p_{1}(t_{3}
-
t_{2})\overline{p}_{2}(t_{3} - t_{2}) \nonumber \\
& + & \overline{p}_{1}(t_{2} - t_{1})\overline{p}_{2}(t_{2} -
t_{1})\overline{p}_{1}(t_{3} - t_{2})p_{2}(t_{3} - t_{2}).
\;\;\;\;\;\;
\end{eqnarray}
In particular, if we assume that the two noises have the same
strength ($\gamma_{1} = \gamma_{2}$) and that the time intervals
during which they act are the same $t_{3} - t_{2} = t_{2} - t_{1} =
T$, we obtain the simplified formula:
\begin{equation}
F(T) = 4 p(t)^3 - 5 p(T)^2 + 2 p(T), \;\;  p(T) = \frac{1 +
e^{-2\gamma T}}{2},
\end{equation}
which shows, as expected, that $F$ starts from 1 and decreases
exponentially in time to $1/4$: the formula displays the whole time
evolution.

\section{Application 2: Transfer of an entangled state through a spin
chain.}
\label{sec:four}

\begin{figure*}[t]
\begin{center}
\includegraphics[scale=0.9]{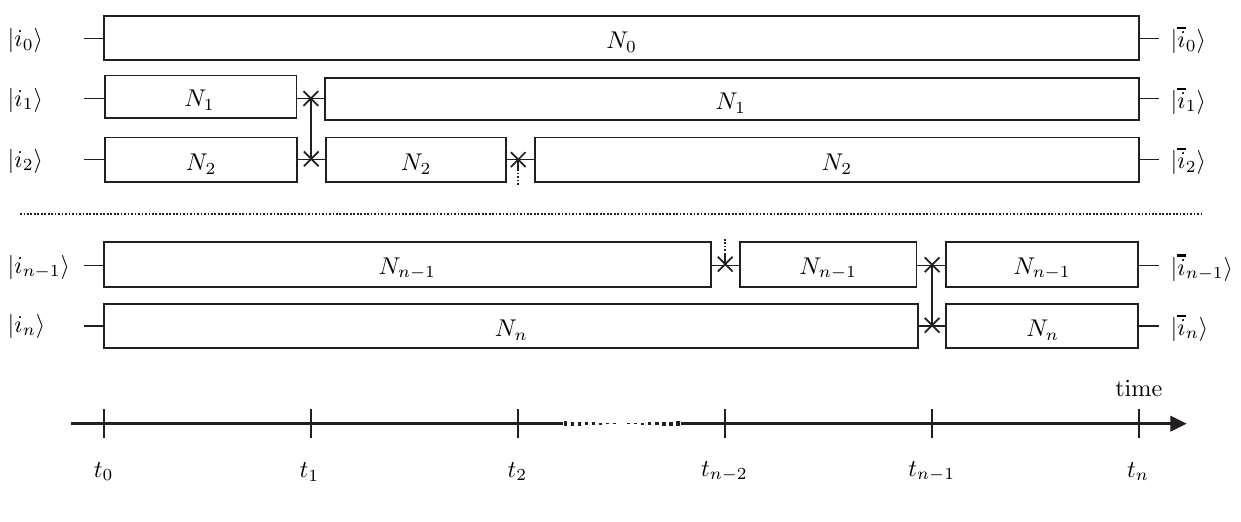}
\caption{Quantum circuit for the transmission of an entangled state
through a spin chain in a noisy environment.} \label{SpCh}
\end{center}
\end{figure*}

One of the most common problems in quantum information theory is the
transfer of information through a noisy channel, which is often
analyzed by modeling the channel with a spin chain~\cite{sc}. By means of the spin chain (see Fig.~\ref{SpCh}) we want to demonstrate the power of the noise gate formalism for analyzing whole quantum circuits.
For the spin chain we chose a standard model consisting of a  chain
of $n+1$ qubits, such that the first two qubits are in a given
normalized entangled state $|\psi\rangle$, while the remaining
qubits are in a general normalized state $|\phi\rangle$; then the
global initial state is $|\Psi\rangle = |\psi\rangle \otimes
|\phi\rangle$ with
\begin{eqnarray} \label{eq:dfdfgfx}
|\psi\rangle & = & \sum_{i_{0},i_{1} = 0}^{1} a_{i_{0},i_{1}}
|i_{0}i_{1}\rangle \\
|\phi\rangle & = & \sum_{i_{2}, \ldots i_{n} = 0}^{1} c_{i_{2},
\ldots i_{n}} |i_{2} \ldots i_{n}\rangle.
\end{eqnarray}
The state of qubit 1 can be transferred to the far end of the chain
by means of a sequence of $n-2$ swap operations; for a circuit not
subject to noise, at the end of the protocol the first and last
qubits are in the entangled state $|\psi\rangle$, factorized from the
rest of the chain: we now analyze how the protocol is changed by the
effect of a noisy environment, which we describe by $n+1$ noise
gates $N_{k}(t_{i},t_{j})$, $k = 1, \ldots, n$, each acting on a
different qubit for the time interval ($t_{j}, t_{i}$). We assume
that the environment is random enough to act independently on each
qubit; this means that we assume that the noise gates are
independent.

The time evolution of the global state can now be immediately
computed: due to the {\it linearity} of the noise gates, and with
reference to Fig.~\ref{SpCh}, one easily gets for the final state at
time $t = t_{n}$
\begin{equation}
| \overline{\Psi} \rangle = \sum_{i_{0}, \ldots i_{n}}
a_{i_{0},i_{1}} \, c_{i_{2}, \ldots i_{n}}\, |\bar{i}_{0}\rangle
\otimes|\bar{i}_{1}\rangle \otimes \ldots |\bar{i}_{n}\rangle,
\end{equation}
where the states $|\bar{i}_{k}\rangle$ are defined as follows:
\begin{eqnarray}
|\bar{i}_{0}\rangle & = & N_{0}(t_n,
t_{0})\, |i_{0}\rangle,  \nonumber \\
|\bar{i}_{k}\rangle & = & N_{k}(t_n, t_{k})
N_{k+1}(t_{k},
t_{0})\, |i_{k+1}\rangle, \;\; k = 1, \ldots n-1 \nonumber \\
|\bar{i}_{n}\rangle & = & \prod_{k=0}^{n-1} N_{n-k}(t_{n-k},
t_{n-k-1})\, |i_{1}\rangle; \label{eq:ev2}
\end{eqnarray}
so the problem is mathematically solved. In order to test the effect of
the environment on the transmission protocol, we compute the reduced
density matrix referring to the 0th and $n$th qubits, obtained from
the full density matrix ${\mathbb E} [ | \overline{\Psi} \rangle
\langle \overline{\Psi} | ]$ by tracing away all other degrees of
freedom (from 1 to $n-1$):
\begin{eqnarray}
\rho_{(0,n)}(t_{n}) & = & \text{Tr}_{1,\ldots, (n-1)} {\mathbb E} [
| \overline{\Psi} \rangle \langle
\overline{\Psi} | ] \\
& = & {\mathbb E} [ |\bar{\psi}\rangle\langle \bar{\psi}| \cdot
\text{Tr} | \bar{\phi} \rangle \langle
\bar{\phi}|],
\end{eqnarray}
where we have defined:
\begin{eqnarray}
|\bar{\psi}\rangle & = &  \sum_{i_{0},i_{1} = 0}^{1} a_{i_{0},i_{1}}
N_{0}(t_{n}, t_{0})\, |i_{0}\rangle
\overline{N}\, |i_{1}\rangle \nonumber \label{eq:fg} \\
| \bar{\phi} \rangle & = &  \sum_{i_{2}, \ldots i_{n} = 0}^{1}
c_{i_{2}, \ldots i_{n}} \underline{N}_1\, |i_{2}\rangle \ldots
\underline{N}_{n-1}\, |i_{n}\rangle, \label{eq:fg2}
\end{eqnarray}
and we have used the short-hand notation:
\begin{eqnarray}
\underline{N}_k & \equiv & N_{k}(t_n, t_{k}) N_{k+1}(t_{k}, t_{0})
\\
\overline{N} & \equiv & \prod_{k=0}^{n-1} N_{n-k}(t_{n-k},
t_{n-k-1}).\label{eq:barN}
\end{eqnarray}
Note that the state $|\bar{\psi}\rangle$ is a linear combination of
the two states $|\bar{i}_{0}\rangle$ and $|\bar{i}_{n}\rangle$ (see
Fig.~\ref{SpCh})---this is the reason why the trace does not affect
$|\bar{\psi}\rangle\langle \bar{\psi}|$---while $| \bar{\phi}
\rangle$ is a linear combination of the remaining states
$|\bar{i}_{1}\rangle, \ldots |\bar{i}_{n-1}\rangle$.

In our setting, \textsc{Property 1} together with our assumption of independence
of the environments lead to the independence of the
statistics of the noise gates $N_i(t_v,t_u)$ and $N_j(t_s,t_r)$
whenever $i\neq j$ or $(t_u,t_v)\cap(t_r,t_s)$ is empty. One
easily verifies that $N_0(t_n,t_0)$, $\overline N$ and $\underline
N_k$ for $k=1,\ldots,n-1$ are independent; this can be immediately checked in Fig.~\ref{SpCh}. This also means that the statistics of
$|\bar{\psi}\rangle\langle \bar{\psi}|$ is independent of that of
$| \bar{\phi} \rangle \langle \bar{\phi}|$, and their average values
can be computed separately:
\begin{eqnarray}
\rho_{(0,n)}(t_{n}) & = & {\mathbb E} [ |\bar{\psi}\rangle\langle \bar{\psi}| ] \cdot {\mathbb E} [
\text{Tr} | \bar{\phi} \rangle \langle
\bar{\phi}|].
\end{eqnarray}
Another important property of the noise gates,  which we shall now
use to simplify the above formula, is the following.

\vskip 0.3cm
\noindent \textsc{Property 2.} Given an initial normalized $n$-qubit
state $|\psi_{0}\rangle$ which evolves, according to a SDE of the
type~\eqref{eq:lsd}, to a random state $|\psi_{t}\rangle =
N(t,t_{0})|\psi_{0}\rangle$, then the following equality holds true:
$\text{Tr}\, {\mathbb E} [ |\psi_{t} \rangle \langle \psi_{t} | ] =
1$ for any $t$. This property is a direct consequence of
Eq.~\eqref{eq:relgfg} and of the fact that $\rho(t)$ satisfies
Eq~\eqref{eq:master}, which is of the Lindblad type and thus trace
preserving.
\vskip 0.3cm

\noindent For the sake of brevity, we denote $i_2,\ldots,i_n$ by
$\underline{i}$; the full expression of $\text{Tr} \mathbb E[
|\ket{\bar\phi}\bra{\bar\phi}]$ is, according to Eq.~\eqref{eq:fg2},
\begin{widetext}
\begin{align}
\text{Tr} \mathbb E[ |\ket{\bar\phi}\bra{\bar\phi}] =
\sum_{\underline{j}} \mathbb E\Big[
\sum_{\underline{i},\underline{i}'} c_{\underline{i}}
c^*_{\underline{i}'} \prod_{k=2}^n
\braket{j_k|N_{k-1}(t_n,t_{k-1})N_{k}(t_{k-1},t_0)|i_k}
\braket{i'_k|N^*_{k}(t_{k-1},t_0)N^*_{k-1}(t_n,t_{k-1})|j_k} \Big];
\end{align}
we now insert two identities $\sum_{\underline{l}}
\ket{\underline{l}} \bra{\underline{l}}$ between the noise matrices,
and after a rearrangement of the terms we get
\begin{align} \label{eq:dfgdf}
\text{Tr} \mathbb E[ |\ket{\bar\phi}\bra{\bar\phi}] =
\sum_{\underline{i},\underline{i}',\underline{l},\underline{l}'}
\prod_{k=2}^n \underbrace{\text{Tr} \mathbb E\Big[
N_{k-1}(t_n,t_{k-1})\ket{l_k}\bra{l'_k}N^*_{k-1}(t_n,t_{k-1})
\Big]}_{(\Delta)} \cdot \mathbb E\Big[c_{\underline{i}}
c^*_{\underline{i}'} \braket{l_k|N_{k}(t_{k-1},t_0)|i_k}
\braket{i'_k|N^*_{k}(t_{k-1},t_0)|l'_k} \Big],
\end{align}
\end{widetext}
where the factorization of the two average values is again
justified by the assumption of independence of the environments and by
Property 1.

By Property 2,  we have
\begin{align}
(\Delta) \; = \; \text{Tr} \mathbb E\Big[ \ket{l_k}\bra{l'_k} \Big]
\; = \; \delta_{l_k,l'_k},
\end{align}
so Eq.~\eqref{eq:dfgdf} simplifies as follows:
\begin{eqnarray}
\lefteqn{\text{Tr} \mathbb E[ |\ket{\bar\phi}\bra{\bar\phi}] =} & &
 \\
& = & \text{Tr} \mathbb E\Big[\sum_{\underline{i},\underline{i}'}
\prod_{k=2}^n c_{\underline{i}} c^*_{\underline{i}'}
N_{k}(t_{k-1},t_0)\ket{i_k} \bra{i'_k}N^*_{k}(t_{k-1},t_0) \Big],
\nonumber
\end{eqnarray}
which, again by Property 2, gives
\begin{align}
\text{Tr} \mathbb E[ |\ket{\bar\phi}\bra{\bar\phi}] =
\text{Tr}\ket\phi\bra\phi = 1.
\end{align}
Accordingly we are left with the expected simple result:
\begin{equation}\label{eq:rho}
\rho_{(0,n)}(t_{n}) \; = \; {\mathbb E} [ | \bar{\psi} \rangle
\langle \bar{\psi} | ],
\end{equation}
from which any relevant piece of information can be obtained.

\subsection{Fidelity of the transmission protocol}

As an application of this formula we now compute the {\it fidelity}
\begin{align}\label{eq:fid}
  F = \text{Tr}[\rho_{(0,n)}(t_{n}) |\psi\rangle\langle\psi|] =
{\mathbb E} [ |\langle \psi |\bar{\psi} \rangle |^2]
\end{align}
of the transmission protocol; here, again, the noise gates come in handy in the computation since we may work with random vectors instead of
density matrices. In order to focus our attention only on the
effect of the noise on the qubit that has been transmitted, we
neglect the effect of the noise gate $N_{0}(t_{n}, t_{0})$ on the
0th qubit. We denote the random matrix components of $\bar N$
by
\begin{align}
  \bar N = \begin{pmatrix}
    \bar a & \bar b\\
    \bar c & \bar d
  \end{pmatrix}
\end{align}
and compute
\begin{align}\label{eq:sclp}
\braket{\psi|\bar\psi} &= \sum_{i_0,i_1,j_0,j_1}a^*_{j_0,j_1}a_{i_0,i_1}
\bra{j_0 j_1} \mathbbm 1\otimes \bar N \ket{i_0 i_1}\\
&= A\bar a + B\bar b + B^* \bar c + (1-A)\bar d,\label{eq:psi}
\end{align}
where
\begin{align}
  A :=|a^{\phantom \star}_{0,0}|^2 + |a^{\phantom \star}_{1,0}|^2, &&
  B :=a^{\star}_{0,0} a^{\phantom \star}_{0,1} +
  a^{\star}_{1,0}a^{\phantom \star}_{1,1}.
\end{align}

In order to become more concrete we chose the {\it amplitude damping
gate} for $N_k$, cf. Eq.~\eqref{eq:bfc4},
\begin{align}
  N_k(t_k,t_{k-1}) &= \begin{pmatrix}
    a_k & b_k\\
    c_k & d_k
  \end{pmatrix}\\
  &:= \begin{pmatrix}
    1 & i\sqrt{\gamma_k}\int_{t_{k-1}}^{t_k}
e^{-\frac{\gamma_k}{2} s} dW^{(k)}_{s}\\
    0 & e^{-\frac{\gamma_k}{2}(t_k-t_{k-1})}
  \end{pmatrix}.
\end{align}
The coupling constant $\gamma_k$ represents the strength  of the
interaction of the $k$-th noise $W^{(k)}$ with the $k$-th qubit. By
definition of $\bar N$, cf. Eq. (\ref{eq:barN}), its
matrix components are given by
\begin{align}\label{eq:barb}
  \bar a = 1, \qquad
  \bar c = 0, \qquad \bar d = \prod_{k=1}^n d_k,
\end{align}
while $\bar b = \bar b_n$ (for $n$ qubits) is defined by the
recursive formula  $\bar b_n := b_n+d_n \bar b_{n-1}$, with $\bar
b_{1} = b_{1}$ where $\bar b_1=b_1$. Now we have all we need to compute  the
fidelity of this protocol. Plugging these matrix components into
Eq.~\eqref{eq:psi} we get by Eq.~\eqref{eq:rho}
\begin{align}
  F&=\mathbb E\left[|A\bar a+B\bar b+(1-A)\bar d|^2\right]\\
  &= A^2 + |B|^2\mathbb E\left[|\bar b|^2\right] + (1-A)^2\bar d^2.
\end{align}
In the last step we have used the fact that only $\bar b$ is random
and that only the terms quadratic in $\bar b$ give a non zero
contribution to the expectation value. Using the recursive
definition of $\bar b$, cf. Eq.~\eqref{eq:barb}, and again
collecting only the terms quadratic in the random variables
$b_k$, we compute
\begin{align}
  \mathbb E\left[|\bar b|^2\right] &= \mathbb E\left[|\bar b_n + d_n \bar b_{n-1}|^2\right]\\
    &= \mathbb E\left[|\bar b_n|^2] + d_n^2\mathbb E[|\bar b_{n-1}|^2\right]\\
    &= 1-e^{-\Gamma}, \quad \Gamma = \sum_{\alpha=1}^{n} \gamma_{\alpha}
(t_{\alpha} -t_{\alpha-1})
\end{align}
by induction. Together with $\bar d=e^{-\frac{\Gamma}{2}}$ we arrive at the formula:
\begin{equation}\label{eq:Famp}
  F_{\makebox{\tiny AmDa}}=\left[A+(1-A)e^{-\frac{\Gamma}{2}}\right]^2 + |B|^2(1-e^{-\Gamma})
\end{equation}

For the \emph{bit flip}, the \emph{phase flip} and the \emph{bit-phase flip} channels,
cf. (\ref{eq:bfc1})-(\ref{eq:bfc3}), we use the noise gates
\begin{align}
    N_l(t_l,t_{l-1}) = \exp\left(i\sqrt{\gamma_l}\sigma_\kappa(W_{t_l}- W_{t_{l-1}})\right)
\end{align}
with $\kappa=x,z,y$, respectively, and so compute $\bar N$ for the three cases according to equation (\ref{eq:barN}):
\begin{align}
  \bar N_{\makebox{\tiny BitFl}} =  \begin{pmatrix}
    \cos \phi && i\sin\phi\\
    i\sin\phi && \cos\phi
  \end{pmatrix},
\end{align}
\begin{align}
  \bar N_{\makebox{\tiny PhFl}} = \begin{pmatrix}
    e^{i \phi} && 0\\
    0 && e^{-i \phi}
  \end{pmatrix},
\end{align}
\begin{align}
  \bar N_{\makebox{\tiny BitPhFl}} = \begin{pmatrix}
    \cos \phi && \sin\phi\\
    -\sin\phi && \cos\phi
  \end{pmatrix}
\end{align}
for
\begin{align}
  \phi=\sum_{k=0}^{n-1} \sqrt{\gamma_k}\left(W_{t_{n-k}}-W_{t_{n-k-1}}\right).
\end{align}
Using equation (\ref{eq:sclp}) the fidelity turns out to be
\begin{align}
  F_{\makebox{\tiny BitFl}} &= \mathbb E\left(\cos^2\phi+4\text{Re} B^2\sin^2\phi\right)\\
  F_{\makebox{\tiny PhFl}} &= \mathbb E\left(|1+A(e^{2i\phi}-1)|^2\right)\\
  F_{\makebox{\tiny BitPhFl}} &= \mathbb E\left(\cos^2\phi+4\text{Im} B^2\sin^2\phi\right),
\end{align}
Here $\text{Re}$ and $\text{Im}$ denote the real and imaginary parts respectively. After evaluation of the expectation value the fidelity is given by
\begin{equation}\label{eq:Fk}
F_{\kappa} =
\frac{1 + g_{\kappa}^2}{2} + \frac{1 - g_{\kappa}^2}{2} e^{-2 \Gamma},
\end{equation}
where $\kappa$ denotes $\makebox{\tiny BitFl}$, $\makebox{\tiny PhFl}$ $\makebox{\tiny BitPhFl}$, and $g_{\makebox{\tiny
BitFl}} = 2 \text{Re}\,B$, $g_{\makebox{\tiny PhFl}}
= 2A - 1$ and $g_{\makebox{\tiny Bit-PhFl}} = 2 \text{Im}\, B$.
As expected, the fidelity decreases exponentially in
time, reaching an asymptotic finite value which depends both on the
initial entangled state and on the type of noise. More generally,
the above formula displays the full dependence of $F$ on the
different parameters entering the protocol, in particular on the
time between two subsequent application of a swap operation and on the
strength of the different noises. It then applies, e.g. to non homogeneous environments, where some qubits feel a stronger
decoherence effect than others. One can easily generalize the
above result by including also uncertainties in the times at which
the different swap operations are applied.

\section{Linear Combinations of Noises}
\label{sec:five}

So far we have looked only at SDEs that are explicitly solvable. In this section we want to consider more complicated noise channels for which an explicit solution might not be available. In the chosen examples of this section we shall see that this is already the case when we combine two or more of the noise channels that we have discussed so far.

In general, a linear combination of noises acting on the same quantum system can be described by
\begin{equation} \label{eq:lsd2}
d |\psi_{t} \rangle = \sum_{\kappa} \left[ i
\sqrt{\gamma^{(\kappa)}}\, L_\kappa\, dW_{t}^{(\kappa)} - \frac{1}{2}\, \gamma^{(\kappa)} L_\kappa^{\dagger} L_\kappa dt \right] |\psi_{t} \rangle,
\end{equation}
where in contrast to (\ref{eq:lsd}) we have neglected the Hamiltonian but in addition have multiple Brownian motions $W_t^{(\kappa)}$. For Lindblad operators $L_\kappa$ this in turn leads to the corresponding master equation
\begin{equation} \label{eq:master2}
\frac{d}{dt}\, \rho(t) =  \sum_\kappa \left[\gamma^{(\kappa)} L_\kappa \rho(t) L^{\dagger}_\kappa - \frac{\gamma^{(\kappa)}}{2} \, \{
L^{\dagger}_\kappa L_\kappa, \rho(t) \}\right]
\end{equation}
in the sense that (\ref{eq:relgfg}) holds accordingly.

As an important feature of the noise gate formalism we notice that, since quantum averages are always expressed as the square modulus of the scalar product of two vectors, the coefficients of the noise gates always enter the
stochastic averages in quadratic combinations, which with a little abuse of terminology we shall refer to as second moments. Now, although (\ref{eq:lsd2}) might not in general be explicitly solvable it is often possible to infer from it all second moments of the noise gate that it describes either analytically or numerically; see~\cite{arn} for a discussion of this topic. As we shall demonstrate, this can be done in an easy way whenever a solution to the corresponding master equation is available. Having these second moments computed either analytically or numerically, one may still work in the noise gate picture even without having the explicit form of the noise gate, which in many circumstances can be more intuitive and faster. For the following discussion let us denote the $(i,j)$-th unknown coefficient of a noise gate $N(t_b,t_a)$ by $n^{(ij)}(t_b,t_a)$, $i,j=0,1$. Then, if the second moments of this noise gate, i.e. $\mathbb E\left(n^{(ij)}(t_b,t_a){n^{(kl)}(t_b,t_a)}^*\right)$ for any $i,j,k,l=0,1$, are available one may perform any computation of a quantum average using the noise gate formalism and in the end plug in the second moments when evaluating the stochastic average.

In order to compute the second moments whenever a solution of the master equation is available, consider $\ket{\psi(t_a)}$ to be the initial value of the SDE and $\rho(t_a)=\ket{\psi(t_a)}\bra{\psi(t_a)}$ the initial value of the master equation, both at time $t_a$. As discussed before, the solution to the SDE at time $t_b$ can be expressed via the noise gate it describes as $N(t_b,t_a)\ket{\psi(t_a)}$. By (\ref{eq:relgfg}) the two entities $\rho(t_b)$ and $\mathbb E(N(t_b,t_a)\ket{\psi(t_a)}\bra{\psi(t_a)}N(t_b,t_a)^*)$ must be equal.
\begin{widetext}
For $\ket{\psi(t_a)}=\sum_i a_i \ket i$ we have
\begin{align}
  \rho_{00}(t_b)&=
    |a_0|^2\mathbb E\left(|n^{(00)}_{T}|^2\right) + |a_1|^2\mathbb E\left(|n^{(01)}_{T}|^2\right) + 2\text{Re} a_0 a_1^*\mathbb E\left(n^{(00)}_{T}{n^{(01)}_{T}}^*\right)\label{eq:r1}\\
  \rho_{01}(t_b)&=
    |a_0|^2\mathbb E\left(n^{(00)}_{T}{n^{(10)}_{T}}^*\right) + |a_1|^2\mathbb E\left(n^{(01)}_{T}{n^{(11)}_{T}}^*\right) +
    a_0 a_1^*\mathbb E\left(n^{(00)}_{T}{n^{(11)}_{T}}^*\right) + a_0^*a_1 \mathbb E\left(n^{(01)}_{T}{n^{(10)}_{T}}^*\right) = \rho_{10}(t_b)^*\label{eq:r2}\\
  \rho_{11}(t_b)&=
    |a_0|^2\mathbb E\left(|n^{(10)}_{T}|^2\right) + |a_1|^2\mathbb E\left(|n^{(11)}_{T}|^2\right) + 2\text{Re} a_0 a_1^*\mathbb E\left(n^{(10)}_{T}{n^{(11)}_{T}}^*\right)\label{eq:r3},
\end{align}
where $\rho_{ij}(t)=\braket{i|\rho(t)|j}$ and $T=t_b-t_a$. Coefficient comparison then easily leads to the second moments.
\end{widetext}

In the following we apply this scheme to the spin chain of the previous section treating two prominent representatives of combined noise channels which are known as \emph{depolarizing} and \emph{generalized amplitude damping} channel, see \cite{nc}.

\subsection{Depolarizing Channel}

The \emph{depolarizing channel} is a linear combination of the bit flip, phase flip and bit-phase flip channels. In terms of Eq. (\ref{eq:lsd2}), $\kappa=1,2,3$ and the $L_\kappa$ are the Pauli matrices $\sigma_x,\sigma_y,\sigma_z$. Here the effect of the environment is to randomly rotate the qubit around the $x,y,z$
axis, with the randomization being proportional to the strength of the
coupling constants $\gamma^{(1)},\gamma^{(2)},\gamma^{(3)}$.

\begin{figure}[b]
\begin{center}
\includegraphics[scale=1]{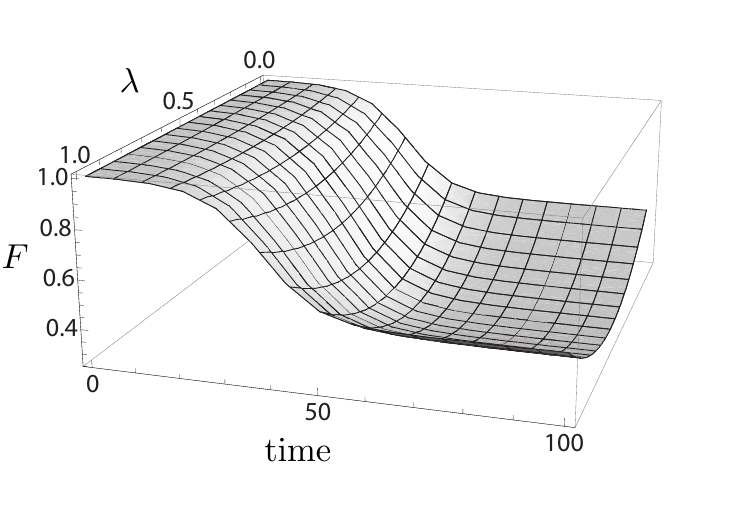}
\caption{Time dependence of the fidelity $F$ of the qubit pair $\sqrt{\lambda}\ket{01}+\sqrt{1-\lambda}\ket{10}$ separated by the described spin chain circuit (see figure (\ref{SpCh})) consisting of $100$ qubits under the influence of depolarizing channels, see Eq. (\ref{eq:depo}). The time intervals between the swap operations were chosen to be equal to $1$ while the coupling coefficients $\gamma^{(i)}_\alpha$ were chosen according to a Gaussian distribution centered around the $50$-th qubit with covariance $\operatorname{cov}_i$ such that $\operatorname{cov}_1:\operatorname{cov}_2:\operatorname{cov}_3=1:2:3$. Due to the chosen class of initial states, the curvature of the surface along the time axis is due to the coupling of $\gamma^{(1)}_\alpha$ and $\gamma^{(2)}_\alpha$ while $\gamma^{(3)}_\alpha$ determines the curvature along the $\lambda$ axis.} \label{fig:depo}
\end{center}
\end{figure}

\begin{widetext}
Hence its master equation takes the following form:
\begin{align}
  \frac{d}{dt}\rho(t)=\begin{pmatrix}
    (\rho_{11}(t)-\rho_{00}(t))\gamma^{(1,2)} && \rho_{10}(t)(\gamma^{(1)}-\gamma^{(2)})-\rho_{01}(t)(\gamma^{(1)}+ \gamma^{(2)} +2\gamma^{(3)})\\
     \rho_{01}(t)(\gamma^{(1)}-\gamma^{(2)})-\rho_{10}(t)(\gamma^{(1)} +\gamma^{(2)}+2\gamma^{(3)})&& (\rho_{00}(t)-\rho_{11}(t))\gamma^{(1,2)}
  \end{pmatrix},
\end{align}
for which
\begin{align}
  \rho_{00}(t_b) &= \frac{1}{2}\left(
    |a_0|^2(1+e^{-2T\gamma^{(1,2)}}) + |a_1|^2(1-e^{-2T\gamma^{(1,2)}})\right)\\
  \rho_{01}(t_b) &= \frac{1}{2}\left(a_0a_1^*(e^{-2T\gamma^{(1,3)}}+e^{-2T\gamma^{(2,3)}}) + a_0^*a_1 (e^{-2T\gamma^{(2,3)}}-e^{-2T\gamma^{(1,3)}})\right)= \rho_{10}(t_b)^*\\
  \rho_{11}(t_b) &=\frac{1}{2}\left(
    |a_0|^2(1-e^{-2T\gamma^{(1,2)}}) + |a_1|^2(1+e^{-2T\gamma^{(1,2)}})\right)
\end{align}
is the solution for initial value $\rho(t_a)$, where we have used $T=t_b-t_a$ and $\gamma^{(m,n)}=\gamma^{(m)}+\gamma^{(n)}$.
\end{widetext}
By coefficient comparison with Eqs. (\ref{eq:r1}), (\ref{eq:r2}) and (\ref{eq:r3}) one finds
\begin{align}
& \mathbb E(n^{(ij)}(t_b,t_a)n^{(i'j')}(t_b,t_a)^{\star}) =\nonumber\\
& = \left\{\begin{array}{ll}
                    \frac{1}{2}(1+e^{-2T\gamma^{(1,2)}}) & ,i=k=i'=k'\\
                    \frac{1}{2}(1-e^{-2T\gamma^{(1,2)}}) &, i=k\neq i'=k'\\
                    \frac{1}{2}(e^{-2T\gamma^{(2,3)}} +e^{-2T\gamma^{(1,3)}}) &, i,k,i',k'=0,0,1,1\\
                    \frac{1}{2}(e^{-2T\gamma^{(2,3)}} -e^{-2T\gamma^{(1,3)}}) &, i,k,i',k'=0,1,1,0\\
                    0 &, \text{else}
                \end{array}\right.
\end{align}
In order to apply this noise channel to the spin chain circuit discussed above
we only need to plug these terms into Eq. (\ref{eq:fid}) where we use the same abbreviations as in Eq. (\ref{eq:sclp}). We shall label the coupling coefficients for the $\alpha$-th qubit by $\gamma_\alpha^{(1)}, \gamma_\alpha^{(2)}, \gamma_\alpha^{(3)}$ for all $\alpha=1,\ldots,n$ when evaluating the product in (\ref{eq:barN}). The computation of the fidelity is then straight forward:
\begin{align}\label{eq:depo}
  &F_{\makebox{\tiny DePo}}=\frac{1}{2}(A^2+(1-A)^2)(1+e^{-2\Gamma^{(1,2)}}) +\nonumber\\
  &+ A(1-A)(e^{-2\Gamma^{(2,3)}}+e^{-2\Gamma^{(1,3)}}) + \nonumber\\
  &+|B|^2(1-e^{-2\Gamma^{(1,2))}})+\text{Re}B^2( e^{-2\Gamma^{(2,3))}} - e^{-2\Gamma^{(1,3))}})
\end{align}
where $\Gamma^{(m,n)}=\sum_{\alpha=1}^N
(\gamma^{(m)}_{\alpha}+\gamma^{(n)}_{\alpha})
(t_{\alpha}-t_{\alpha-1})$ and $\gamma^{(i)}_{\alpha}$ denotes the
$i$th coupling constant of the $\alpha$th noise gate in the
circuit. Note that the formula reduces to $F_\kappa$, Eq.
(\ref{eq:Fk}), when all the coupling constants are set to zero
except the ones associated with one Pauli matrix. Figure \ref{fig:depo} displays the time evolution of the fidelity of the spin chain under the influence of the depolarizing channel for a specific class of initial states.

\subsection{Generalized Amplitude Damping Channel}

The \emph{generalized amplitude damping channel} in turn is a linear combination of the amplitude damping channel as defined in the previous sections and the inverse process for which $|1\rangle$ is stable and $|0\rangle$ decays. In terms of Eq. (\ref{eq:lsd2}), $\kappa=1,2$ and the $L_\kappa$ are the operators $\sigma_{-}$ and $\sigma_{+}=|1\rangle\langle 0|$. The amplitude damping channel that we have discussed in the previous sections is the zero temperature limit of this channel. For non zero temperature the qubit may now also gain energy at the rate $\gamma^{(2)}$.

\begin{figure}[b]
\begin{center}
\includegraphics[scale=1]{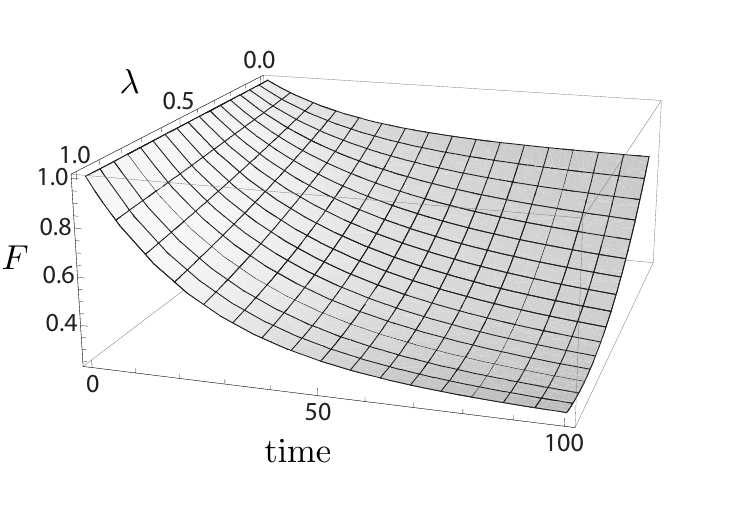}
\caption{Time dependence of the fidelity $F$ of the qubit pair $\sqrt{\lambda}\ket{01}+\sqrt{1-\lambda}\ket{10}$ separated by the described spin chain circuit (see figure (\ref{SpCh}))  consisting of $100$ qubits under the influence of generalized amplitude damping channels, see Eq. (\ref{eq:fidampdamp}). The time intervals between the swap operations were chosen to be equal $1$ while $\gamma^{(1)}:\gamma^{(2)}:(\gamma^{(1)}+\gamma^{(2)})=3:1:4$. The latter is reflected in the curvature along the $\lambda$ axis.} \label{fig:ampdamp}
\end{center}
\end{figure}

\begin{widetext}
This time its master equation takes the following form:
\begin{align}
  \frac{d}{dt}\rho(t)=\begin{pmatrix}
    \rho_{11}(t)\gamma^{(1)}-\rho_{00}(t)\gamma^{(2)} && -\frac{\rho_{01}(t)}{2}(\gamma^{(1)}+\gamma^{(2)})\\
     -\frac{\rho_{10}(t)}{2}(\gamma^{(1)}+\gamma^{(2)}) && \rho_{00}(t)\gamma^{(2)}-\rho_{11}(t)\gamma^{(1)}
  \end{pmatrix},
\end{align}
for which
\begin{align}
  \rho(t_b) = \begin{pmatrix}
    |a_0|^2 \cdot \frac{\gamma^{(1)}+\gamma^{(2)}e^{-\Gamma T}}{\Gamma} + |a_1|^2 \cdot \frac{\gamma^{(1)}(1-e^{-\Gamma T})}{\Gamma} && a_0a_1^* e^{-\frac{\Gamma}{2}T}\\
    a_0^*a_1 e^{-\frac{\Gamma}{2}T} && |a_0|^2 \cdot \frac{\gamma^{(2)}(1-e^{-\Gamma T})}{\Gamma} + |a_1|^2 \cdot \frac{\gamma^{(2)}+\gamma^{(1)}e^{-\Gamma T}}{\Gamma}
  \end{pmatrix}
\end{align}
is the solution for initial value $\rho(t_a)$, where we used $T=t_b-t_a$ and $\Gamma=\gamma^{(1)}+\gamma^{(2)}$.
\end{widetext}
Again, by coefficient comparison with Eq. (\ref{eq:r1}), (\ref{eq:r2}) and (\ref{eq:r3}) one finds
\begin{align}
 \mathbb E(|n^{(00)}(t_b,t_a)|^2) &= \frac{\gamma^{(1)}+\gamma^{(2)}e^{-\Gamma T}}{\Gamma}\\
 \mathbb E(|n^{(01)}(t_b,t_a)|^2) &= \frac{\gamma^{(1)}}{\Gamma}e^{-\Gamma T}\\
 \mathbb E(|n^{(10)}(t_b,t_a)|^2) &= \frac{\gamma^{(2)}}{\Gamma}e^{-\Gamma T}\\
 \mathbb E(|n^{(11)}(t_b,t_a)|^2) &= \frac{\gamma^{(2)}+\gamma^{(1)}e^{-\Gamma T}}{\Gamma}
\end{align}
and
\begin{align}
 \mathbb E(n^{(00)}(t_b,t_a){n^{(11)}(t_b,t_a)}^*) &= e^{-\frac{\Gamma}{2}T}
\end{align}
while all other second moments are equal to zero. As we have done with the depolarizing channel we apply this noise to the spin chain circuit discussed above and therefore we, again, only need to plug these terms into Eq. (\ref{eq:fid}) using the same abbreviations as in Eq. (\ref{eq:sclp}). In order to keep the displayed formulas short we choose the coupling constants to be the same for all qubits, i.e. $\gamma^{(1)}_\alpha=\gamma^{(1)}$ and $\gamma^{(2)}_\alpha=\gamma^{(2)}$ for all $\alpha=1,\ldots,n$, when computing the product in (\ref{eq:barN}). We then find
\begin{align}\label{eq:fidampdamp}
  &F_{\makebox{\tiny GeAmDa}}=(A^2\frac{\gamma^{(1)}}{\Gamma}+(1-A)^2\frac{\gamma^{(2)}} {\Gamma}+|B|^2) + \nonumber\\
  & + (A^2\frac{\gamma^{(2)}}{\Gamma} + (1-A)^2\frac{\gamma^{(1)}}{\Gamma} - |B|^2)e^{-\Gamma(t_n-t_0)} +\nonumber\\
  &+2A(1-A)e^{-\frac{\Gamma}{2}(t_n-t_0)}
\end{align}
for $\Gamma=\gamma^{(1)}+\gamma^{(2)}$ and
$\gamma^{(i)}=\gamma^{(i)}_\alpha$,  such that all noise gates
$N_{\alpha}$ in the circuit have the same coupling constants. Note
that also this formula reduces to (\ref{eq:Famp}) if
$\gamma^{(2)}$ is set to zero. An example of the time evolution of the fidelity for a specific class of initial conditions under the influence of generalized amplitude damping is shown in figure \ref{fig:ampdamp}.

\section{Conclusion}
\label{sec:six}

We have suggested the noise gate formalism as a handy
approach for analyzing the effect of the environment on quantum
algorithms; it is very intuitive as it allows the influence
of the environment to be treated in terms of noise gates, which can be manipulated
like any other quantum gate. In many situations it makes the
computation easier, either analytically or numerically. We emphasize
again that it is especially interesting for numerical simulations
because linear SDEs can be integrated by standard methods
\cite{klo}. In contrast to solving the Lindblad equation
numerically, which roughly scales quadratically with the number of
degrees of freedom, the numerical integration of (\ref{eq:lsd})
scales only linearly, even if the noises are dependent. Finally note
that SDEs can also be generalized to model non Markovian quantum
noise.

\noindent \textsc{Acknowledgements.} The work of A.B. has been partly
supported by DFG (Germany). The work of D.-A.D. has been supported by
the EU Grant No. ERG\;044941-STOCH-EQ.


\end{document}